\documentclass[conference]{IEEEtran} 
\IEEEoverridecommandlockouts
\usepackage{cite}
\usepackage{amsmath,amssymb,amsfonts}
\usepackage{graphicx}
\usepackage{textcomp}
\usepackage{xcolor}
\graphicspath{{./images/}} 
\usepackage{wrapfig}       
\usepackage{caption}       
\usepackage{subcaption}    
\usepackage[hidelinks]{hyperref} 

\usepackage[all]{hypcap}
\usepackage[nameinlink,capitalise]{cleveref} 
\def\BibTeX{{\rm B\kern-.05em{\sc i\kern-.025em b}\kern-.08em
    T\kern-.1667em\lower.7ex\hbox{E}\kern-.125emX}}

\usepackage[linesnumbered,ruled,vlined]{algorithm2e}
\usepackage{algpseudocode}
\usepackage{siunitx} 

\graphicspath{{./}} 
\usepackage{booktabs}
\usepackage{multirow}
\usepackage{tikz}          
\usetikzlibrary{arrows}    
\usepackage{listings}
\usepackage{tcolorbox}
\usepackage{colortbl}
\usepackage{amssymb}
\usepackage{pifont}

\usepackage{paralist} 

\usepackage{color}
\usepackage{soul}
\definecolor{hl}{rgb}{0.80,0.80,0.80}
\sethlcolor{hl}
\definecolor{highlight}{rgb}{0.9,1.0,0.9}

\usepackage[all]{hypcap}
\usepackage[nameinlink,capitalise]{cleveref} 

\usepackage{titlesec}
	\titlespacing*{\section}{0pt}{*0.8}{*0.6}
	\titlespacing*{\subsection}{0pt}{*0.6}{*0.4}
	\titlespacing*{\subsubsection}{0pt}{*0.4}{*0.2}

\begin{document}

\title{The Last Dependency Crusade: Solving Python Dependency Conflicts with LLMs}
\author{\IEEEauthorblockN{Antony Bartlett}
\IEEEauthorblockA{\textit{Delft University of Technology}\\
Delft, The Netherlands \\
a.j.bartlett@tudelft.nl}
\and
\IEEEauthorblockN{Cynthia Liem}
\IEEEauthorblockA{\textit{Delft University of Technology}\\
Delft, The Netherlands \\
c.c.s.liem@tudelft.nl}
\and
\IEEEauthorblockN{Annibale Panichella}
\IEEEauthorblockA{\textit{Delft University of Technology}\\
Delft, The Netherlands \\
a.panichella@tudelft.nl}
}
\maketitle

\newcommand{\eg}{\textit{e.g.,}}
\newcommand{\Eg}{\textit{E.g.,}}
\newcommand{\ie}{\textit{i.e.,}}
\newcommand{\Ie}{\textit{I.e.,}}
\newcommand{\etal}{\textit{et al.}}
\newcommand{\etc}{\textit{etc.}}
\newcommand{\wrt}{\textit{w.r.t.}}
\newcommand{\cfr}{\textit{cfr.}}
\newcommand{\Cfr}{\textit{Cfr.}}
\newcommand{\viz}{\textit{viz.}}
\newcommand{\aka}{\textit{a.k.a.}}
\newcommand{\cf}{\textit{cf.}}
\newcommand{\Cf}{\textit{Cf.}}
\newcommand{\approach}{\texttt{PLLM}}
\newcommand{\baseline}{\texttt{PyEGo}}
\newcommand{\baselinetwo}{\texttt{ReadPyE}}
\newcommand{\hgdataset}{\texttt{HG2.9K}}

\newcommand{\pvalue}{p\text{-value}}
\newcommand{\atwelve}{\mathrm{\hat{A}}_{12}}

\newtheorem{mydef}{Definition}
\newtheorem{probdef}{Problem Definition}
\captionsetup{
  justification = centering
}
\tikzstyle{mybox} = [draw=black, thick, rectangle, rounded corners, inner ysep=5pt, inner xsep=5pt] 

\colorlet{punct}{red!60!black}
\definecolor{background}{HTML}{EEEEEE}
\definecolor{delim}{RGB}{20,105,176}
\colorlet{numb}{magenta!60!black}

\lstdefinelanguage{json}{
    basicstyle=\normalfont\ttfamily,
    numbers=left,
    numberstyle=\scriptsize,
    stepnumber=1,
    numbersep=8pt,
    showstringspaces=false,
    breaklines=true,
    frameround=fttt,
    frame=trBL,
    backgroundcolor=\color{background},
    captionpos=b,
    literate=
     *{0}{{{\color{numb}0}}}{1}
      {1}{{{\color{numb}1}}}{1}
      {2}{{{\color{numb}2}}}{1}
      {3}{{{\color{numb}3}}}{1}
      {4}{{{\color{numb}4}}}{1}
      {5}{{{\color{numb}5}}}{1}
      {6}{{{\color{numb}6}}}{1}
      {7}{{{\color{numb}7}}}{1}
      {8}{{{\color{numb}8}}}{1}
      {9}{{{\color{numb}9}}}{1}
      {:}{{{\color{punct}{:}}}}{1}
      {,}{{{\color{punct}{,}}}}{1}
      {\{}{{{\color{delim}{\{}}}}{1}
      {\}}{{{\color{delim}{\}}}}}{1}
      {[}{{{\color{delim}{[}}}}{1}
      {]}{{{\color{delim}{]}}}}{1},
}

\lstdefinelanguage{Python}{
    basicstyle=\normalfont\ttfamily,
    numbers=left,
    numberstyle=\scriptsize,
    stepnumber=1,
    numbersep=8pt,
    showstringspaces=false,
    breaklines=true,
    frameround=fttt,
    frame=trBL,
    backgroundcolor=\color{background},
    captionpos=b,
    keywordstyle=\color{blue}\bfseries,
    keywordstyle={[2]\color{orange}\bfseries}, 
    commentstyle=\color{gray}\itshape,
    stringstyle=\color{red},
    identifierstyle=\color{black},
    keywords={%
        and, as, assert, break, class, continue, def, del, elif, else, except, PromptTemplate, chain, %
        False, finally, for, from, global, if, import, in, is, lambda, None, %
        nonlocal, not, or, pass, raise, return, True, try, while, with, yield%
    },
    morekeywords=[2]{prompt, template, partial_variables, llm, parser}, 
    emphstyle=\color{blue}\bfseries,
    emph={self},
    literate=
      {0}{{{\color{numb}0}}}{1}
      {1}{{{\color{numb}1}}}{1}
      {2}{{{\color{numb}2}}}{1}
      {3}{{{\color{numb}3}}}{1}
      {4}{{{\color{numb}4}}}{1}
      {5}{{{\color{numb}5}}}{1}
      {6}{{{\color{numb}6}}}{1}
      {7}{{{\color{numb}7}}}{1}
      {8}{{{\color{numb}8}}}{1}
      {9}{{{\color{numb}9}}}{1}
      {=}{{{\color{punct}{=}}}}{1}
      {+}{{{\color{punct}{+}}}}{1}
      {-}{{{\color{punct}{-}}}}{1}
      {*}{{{\color{punct}{*}}}}{1}
      {/}{{{\color{punct}{/}}}}{1}
      {<}{{{\color{punct}{<}}}}{1}
      {>}{{{\color{punct}{>}}}}{1}
      {,}{{{\color{punct}{,}}}}{1}
      {;}{{{\color{punct}{;}}}}{1}
      {)}{{{\color{punct}{)}}}}{1}
      {(}{{{\color{punct}{(}}}}{1}
      {:}{{{\color{punct}{:}}}}{1}
      {.}{{{\color{punct}{.}}}}{1},
}

\definecolor{promptcolor}{rgb}{0.0, 0.0, 0.5}
\definecolor{promptbgcolor}{rgb}{0.9, 0.9, 0.9}
\definecolor{variablecolor}{rgb}{0.5, 0.0, 0.5}

\newcommand{\highlightvariable}[1]{\textcolor{variablecolor}{\{#1\}}}

\lstdefinestyle{python}{
    basicstyle=\ttfamily\scriptsize,
    backgroundcolor=\color{background},
    frameround=fttt,
    frame=trBL,
    breaklines=true,
    postbreak=\mbox{\textcolor{red}{$\hookrightarrow$}\space},
    numbers=none,
    showstringspaces=false,
    keywordstyle=\color{blue}\bfseries,
    keywordstyle={[2]\color{orange}\bfseries}, 
    keywordstyle={[3]\color{black}\bfseries}, 
    keywords={import},
    morekeywords=[2]{sklearn},
    morekeywords=[3]{scikit, -, learn},
    commentstyle=\color{promptcolor},
    captionpos=b,
}

\lstdefinestyle{llmprompt}{
    basicstyle=\ttfamily\scriptsize,
    backgroundcolor=\color{white},
    frameround=fttt,
    frame=trBL,
    breaklines=true,
    postbreak=\mbox{\textcolor{red}{$\hookrightarrow$}\space},
    numbers=none,
    showstringspaces=false,
    keywordstyle=\color{blue}\bfseries,
    keywordstyle={[2]\color{orange}\bfseries}, 
    keywords={Prompt},
    morekeywords=[2]{error_msg, module_name, module_versions, previous_versions, format_instructions,  raw_file, python_version},
    commentstyle=\color{promptcolor},
    captionpos=b,
}

\definecolor{dockerbg}{RGB}{211, 211, 211}
\definecolor{dockerkey}{rgb}{0.0, 0.0, 0.6}
\definecolor{dockerstring}{rgb}{0.6, 0.0, 0.0}
\definecolor{dockercomment}{rgb}{0.0, 0.5, 0.0}

\lstdefinelanguage{Dockerfile}{
	keywords={FROM, MAINTAINER, RUN, CMD, LABEL, EXPOSE, ENV, ADD, COPY, ENTRYPOINT, VOLUME, USER, WORKDIR, ARG, ONBUILD, STOPSIGNAL, HEALTHCHECK, SHELL},
	keywordstyle=\color{dockerkey}\bfseries,
	ndkeywords={},
	ndkeywordstyle=\color{dockerkey}\bfseries,
	identifierstyle=\color{black},
	sensitive=false,
	comment=[l]\#,
	commentstyle=\color{dockercomment}\ttfamily,
	stringstyle=\color{dockerstring}\ttfamily,
	morestring=[b]'
}

\lstdefinestyle{docker}{
	basicstyle=\ttfamily\small,
	backgroundcolor=\color{dockerbg},
	frameround=fttt,
    frame=trBL,
	breaklines=true,
	postbreak=\mbox{\textcolor{red}{$\hookrightarrow$}\space},
	numbers=none,
	showstringspaces=false,
	captionpos=b,
}

\begin{abstract}\label{sec:abstract}
Resolving Python dependency issues remains a tedious and error-prone process, forcing developers to manually trial compatible module versions and interpreter configurations. Existing automated solutions, such as knowledge-graph-based and database-driven methods, face limitations due to the variety of dependency error types, large sets of possible module versions, and conflicts among transitive dependencies. This paper investigates the use of Large Language Models (LLMs) to automatically repair dependency issues in Python programs. We propose \approach{} (pronounced “plum”), a novel retrieval-augmented generation (RAG) approach that iteratively infers missing or incorrect dependencies. \approach{} builds a test environment where the LLM proposes module combinations, observes execution feedback, and refines its predictions using natural language processing (NLP) to parse error messages. We evaluate \approach{} on the Gistable \hgdataset{} dataset, a curated collection of real-world Python programs. Using this benchmark, we explore multiple PLLM configurations, including six open-source LLMs evaluated both with and without RAG. Our findings show that RAG consistently improves fix rates, with the best performance achieved by \texttt{Gemma-2 9B} when combined with RAG. Compared to two state-of-the-art baselines, \baseline{} and \baselinetwo{}, \approach{} achieves significantly higher fix rates; +15.97\% more than \baselinetwo{} and +21.58\% more than \baseline{}. Further analysis shows that \approach{} is especially effective for projects with numerous dependencies and those using specialized numerical or machine-learning libraries.
\end{abstract}


\section{Introduction}\label{sec:introduction}

Python, introduced in 1991, has become one of the most widely used programming languages~\cite{ieeetopprogramming}, due in part to its extensive ecosystem of reusable modules. These modules, once imported, become project dependencies that must be managed over time. As Python’s ecosystem grew, so did the complexity of dependency management, prompting the development of tools such as \texttt{distutils}\cite{smith2015cython}, \texttt{setuptools}\cite{setuptools}, and \texttt{pip}~\cite{pip-citation}, which are based on the Python Package Index (\texttt{PyPI}) to install direct and transitive dependencies.

Yet despite these advances, many Python programs still fail to run out-of-the-box~\cite{Yang_2016,gistable2018horton}, typically due to missing, incompatible, or conflicting dependencies. A \textit{dependency conflict} occurs when two or more modules require different and incompatible versions of the same dependency. This can arise directly—when a specified version is unavailable or mismatches another requirement—or transitively, when nested dependencies bring incompatible constraints. Such conflicts often result in runtime errors, broken APIs, or failed installations. This problem is particularly acute in machine learning and scientific computing, where libraries frequently depend on hardware-specific versions (e.g., CUDA) or require tightly coupled version combinations~\cite{Huang:fse2023}. Even minor version updates can introduce breaking changes~\cite{dietrich2019dependency}, highlighting the need for precise and automated dependency resolution.

Existing automated solutions rely on knowledge graphs (\eg, \baseline{}~\cite{yepyego2022}, \baselinetwo{}~\cite{cheng2024readpy}), but these struggle with complex dependencies and require frequent updates.

We introduce \approach{}, an LLM-based method that resolves dependency conflicts through iterative repair. \approach{} combines Retrieval-Augmented Generation with build feedback, addressing hallucinations through concrete error feedback~\cite{sallou2024breaking}.  To explore this, we introduce \approach{} (pronounced “plum”), a novel LLM-based method that automatically resolves dependency issues through an iterative repair process. 

We evaluate \approach{} on the Gistable \hgdataset{}, a curated benchmark of real-world challenging Python programs commonly used to assess dependency resolution methods. This benchmark is widely used in the literature to evaluate the performance of dependency conflict resolution techniques~\cite{horton2019dockerizemeautomaticinferenceenvironment, yepyego2022, cheng2024readpy}.
The following research questions guide our study:
\vspace{-1em}
\begin{enumerate}\label{sec:researchquestions}
\item[\textbf{RQ1}:] \textit{To what extent can current LLMs infer working dependencies from a given Python file?}
\item[\textbf{RQ2}:] \textit{How does \approach{} compare to state-of-the-art knowledge-based dependency resolution techniques?}
\item[\textbf{RQ3}:] \textit{Under what conditions does \approach{} outperform traditional methods?}
\end{enumerate}
\vspace{-1em}

We evaluated \approach{} with six LLMs, observing \text{Gemma-2 9B} with RAG as the most performant. Once evaluated against the entire \hgdataset{}, \approach{} was found to significantly outperform knowledge-graph baselines \baseline{} and \baselinetwo{}, achieving +21.58\% and +15.97\% more fixes respectively.

\approach{} is particularly effective for projects with intricate dependency structures, excelling at complex dependencies such as those for machine learning (e.g., \texttt{tensorflow}) and numerical computing (e.g., \texttt{scipy}), contributing the most unique conflict resolutions overall.

\section{Background and Related Work}
\label{sec:related}


\subsection{Existing Approaches}

Knowledge graphs have been widely used to encode relationships among Python modules. Horton and Parnin~\cite{horton2019dockerizemeautomaticinferenceenvironment} introduced \texttt{DockerizeMe}, which constructs a knowledge graph using \texttt{Libraries.io} data stored in Neo4J to infer required dependencies and resolve transitive dependencies.

Building on this work, \baseline{}~\cite{yepyego2022} expands \texttt{DockerizeMe} with package documentation and release metadata, using 256,000 nodes and 1.9 million relationships.

\baselinetwo{}~\cite{cheng2024readpy} combines naming similarity with optimization algorithms for module matching. By iteratively validating dependency choices and adjusting module selections \baselinetwo{} finds solutions based on validation logs.
Although this method improves matching accuracy, it and \baseline{} require regular updates to maintain the graph's relevance, a common limitation among knowledge-graph-based tools.

In contrast to graph-based methods, Mukherjee~\etal{}~\cite{Mukherjeepydfix2021} proposed \texttt{PyDFix}, which uses regex-based parsing of Python error logs to infer missing dependencies. By analyzing build failures and iteratively patching the environment, \texttt{PyDFix} provides a lightweight, runtime-driven alternative. However, regex-based methods are inherently brittle: even small changes in log formatting can break the parsing logic, limiting the generality and robustness of such approaches.

While these methods have contributed significantly to the field, they suffer from key limitations: knowledge graphs require large databases and frequent updates, while regex methods are brittle to format changes~\cite{zhang2023system, zhu2019tools, messaoudi2018search}.

\begin{figure*}[!t]
\centering
\begin{subfigure}{\columnwidth}
\centering
\makebox[0pt]{\includegraphics[width=1.2\linewidth]{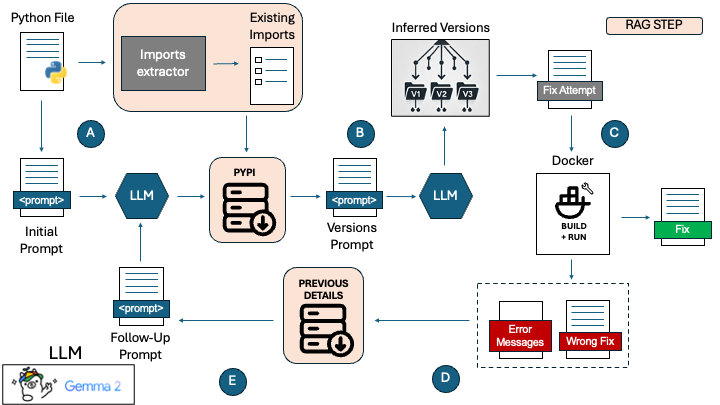}}
\end{subfigure}%
\caption{Overview of \approach{}, which encompasses five main stages:
(A) extracting import statements from the input Python file,
(B) prompting the model to infer modules and Python versions,
(C) generating candidate dependency fixes using PyPI metadata,
(D) validating candidate fixes through Docker build and execution, and
(E) providing error-based feedback to the LLM for iterative refinement.}
\label{fig:process_flow}
\end{figure*}

\approach{} uses an LLM within a RAG pipeline to dynamically resolve dependencies by retrieving PyPI metadata and incorporating build error feedback, avoiding static infrastructure and reducing model hallucination risks~\cite{sallou2024breaking}. This adaptive pipeline reduces the need for pre-built infrastructure and improves generalization across a broad range of dependency error types.

\subsection{Datasets of Python Dependency Conflicts}
Evaluating dependency repair techniques requires datasets that include real-world Python code with known dependency issues. The most widely used benchmark is the Gistable dataset, introduced by Horton and Parnin~\cite{gistable2018horton}. This dataset, collected from GitHub’s Gist platform\footnote{\url{https://gist.github.com/}}, contains over 10,000 Python gists—programs that vary in complexity. The authors found that only 24.4\% of these gists could be executed without modification, while 52\% failed due to missing imports. The Gistable dataset\footnote{\url{https://github.com/gistable/gistable}} includes a challenging subset of 2,891 ``hard'' gists that fail due to missing imports and remain unrunnable without fixes. This subset has become the gold standard for evaluating dependency-fixing approaches.

\section{Our Approach}\label{sec:approach}

\approach{} is a prompt-based LLM system supporting multiple open-source backends. Figure~\ref{fig:process_flow} illustrates our five-stage pipeline combining RAG and LLMs for iterative Python dependency resolution.
\begin{itemize}
\item \textit{Stage A}: Inferring module names and Python version.
\item \textit{Stage B}: Inferring module versions.
\item \textit{Stage C}: Docker-based build and validation.
\item \textit{Stage D}: Error analysis and classification.
\item \textit{Stage E}: Feedback loop and iterative refinement.
\end{itemize}

We now detail each stage, describe the prompts used, and explain how RAG contributes to each step.

\subsection{Inferring Module Names and Python Version (Stage A)}
\label{subsec:stageA}

\approach{} prompts the LLM with the input Python file to infer required modules and Python version (Listing~\ref{lst:promptfile}). Placeholders like \texttt{raw\_file} and \texttt{format\_instructions} (JSON schema in Listing~\ref{lst:schemafile}) make prompts modular and reusable.

When RAG is enabled, \approach{} complements the LLM output by running a regex-based static analysis to extract all import statements from the input. These are merged with the LLM-inferred modules to form a more complete list of module/ dependencies to consider for conflict fixes. We further elaborate on RAG in Section~\ref{subsubsec:rag}.

\begin{lstlisting}[float=t, style=llmprompt, caption={Prompt for inferring information from a given Python file.}, label={lst:promptfile}]
Prompt:
"Given a python file:\n{raw_file}\nReturn a list of Python modules and python version required to run. Output JSON based on the schema {format_instructions}"
\end{lstlisting}

\begin{lstlisting}[float=t, language=json, numbers=none, caption={JSON Schema: Dependencies and Python version.}, label={lst:schemafile}, basicstyle=\scriptsize\ttfamily]
{ "python_modules": [{"module": "<String>", "version": "<String>"}], "python_version": "<String>" }
\end{lstlisting}

\subsection{Inferring Module Versions (Stage B)}
\label{subsec:stageb}
Stage B prompts the LLM to assign versions to each identified module. Using RAG, we provide PyPI metadata to avoid hallucinated versions (Listing~\ref{lst:prompt-version}); without RAG, the LLM infers versions from context (Listing~\ref{lst:prompt-version-ragless}). Standard library modules are filtered out, and import/install name mismatches are resolved via curated mapping, e.g., \texttt{sklearn} $\rightarrow$ \texttt{scikit-learn}, \texttt{bs4} $ \rightarrow$ \texttt{beautifulsoup4} as shown Listing~\ref{lst:sklearn}.

\begin{lstlisting}[float=t, style=llmprompt, caption={Prompt with RAG for selecting a version from available PyPI releases.}, label={lst:prompt-version}]
Prompt:
"Given a comma-separated list of 'Module versions' for the '{module_name}' module, from oldest to newest:\n{module_versions}\nPerform equally distanced sampling to return a version from the given versions, excluding previously used versions ({previous_versions}). Return the information with the format {format_instructions}"
\end{lstlisting}

\begin{lstlisting}[float=t, style=llmprompt, caption={Prompt without RAG (LLM infers version from context).}, label={lst:prompt-version-ragless}]
Prompt:
"Infer a possible working version of the '{module_name}' module for Python {python_version}.\nReturn the information with the format {format_instructions}"
\end{lstlisting}

\begin{lstlisting}[float=t, style=python, caption={Discrepancy between import and install names.}, label={lst:sklearn}]
import sklearn
Must be installed as 'scikit-learn'
\end{lstlisting}

The prompt with RAG in Listing~\ref{lst:prompt-version} contains a larger set of placeholders, as we provide the LLM with specific dependency information for guidance. In the Listing, \texttt{`module\_name'} represents the name of the module (dependency) obtained from the initial prompt (Stage A). When RAG is enabled, we provide \texttt{`module\_versions'}, a comma-separated list of available versions sourced from PyPI, for the given dependency. We also provide the LLM with a list of previously attempted versions \texttt{`previous\_versions'} to prevent using versions that already resulted in unsuccessful conflict fixes. To promote diversity and avoid local optima, we request the LLM to sample versions evenly across the available list. Finally, the \texttt{`format\_instruction'} prompts the LLM to return both the dependency name and the newly chosen version.
When RAG is disabled, Stage B uses a much simpler prompt. An example of such a prompt can be seen in Listing~\ref{lst:prompt-version-ragless}. In this prompt, we only give the LLM the module name, python version and format instructions, allowing it to infer a version only with this information.

\subsection{Docker-based Build and Validation (Stage C)}
\label{subsec:validation}

\approach{} validates predictions by building and executing the Python file in Docker containers. We construct Dockerfiles with the inferred Python version and modules (Listing~\ref{lst:dockerfile}), using unique names for concurrent execution. Success is indicated by a valid configuration; failure triggers error analysis.

\begin{lstlisting}[float=t, language=Dockerfile, style=docker, numbers=none, caption={Dockerfile.}, label={lst:dockerfile}, basicstyle=\scriptsize\ttfamily]
 FROM python:3.6
 WORKDIR /app
 RUN ["pip","install","--upgrade","pip"]
 RUN ["pip","install","--trusted-host","pypi.python.org","--default-timeout=100","keras==2.0.9"]
 RUN ["pip","install","--trusted-host","pypi.python.org","--default-timeout=100","tensorflow==2.4.4"]
 COPY snippet.py /app
 CMD ["python", "/app/snippet.py"]
\end{lstlisting}

\subsection{Error Analysis and Classification (Stage D)}
\label{subsec:error_analysis}
Stage D analyzes build/execution errors to guide LLM corrections. Building on prior work in automated environment inference~\cite{horton2019dockerizemeautomaticinferenceenvironment, cheng2024readpy, yepyego2022}, this hybridized approach reduces over-prompting--the bottleneck of \approach{}--by identifying eight common error types that arise during dependency resolution: \texttt{VersionNotFound}, \texttt{DependencyConflict}, \texttt{ImportError}, \texttt{ModuleNotFound}, \texttt{AttributeError}, \texttt{InvalidVersion}, \texttt{NonZeroCode}, and \texttt{SyntaxError}.

A key insight from our development is that combining multiple failure signals into a single prompt often leads to LLM confusion or hallucinations. Therefore, we adopt a \textit{multi-step prompting strategy} that isolates specific messages and crafts focused prompts, improving interpretability and response accuracy.

For example, \texttt{ImportError}, the most common error, is handled with a specialized prompt to identify the missing module (Listing~\ref{lst:prompt-import-error}).

\begin{lstlisting}[float=t, style=llmprompt, caption={Prompt for identifying the missing module from an ImportError.}, label={lst:prompt-import-error}]
Prompt:
"Given the following ImportError:\n{error_msg}\nIdentify the module causing the error.\nThe module is usually mentioned in a statement like 'from x import y'.\nReturn just the module name using the format {format_instructions}"
\end{lstlisting}

An \texttt{ImportError} indicates a missing module that was not installed. We provide the LLM with the error, requesting it return the offending module name. This ensures the LLM handles noisy logs and correctly extracts the relevant information. Stage D thus acts as the diagnostic core, linking runtime feedback to model-driven correction.

\subsection{Feedback Loop and Iterative Refinement (Stage E)}
\label{subsec:feedback}

Stage E uses error feedback to iteratively refine dependency suggestions. Each cycle attempts to repair by inferring (i) missing modules, (ii) different versions, or (iii) additional dependencies until success or a predefined iteration limit is reached. We maintain a history of previously attempted combinations to prevent redundant retries while continuing to verify the existence and naming of modules on PyPI. This stage is imperative, as it also allows \approach{} to indirectly locate and address transitive dependencies.

\subsection{Retrieval-Augmented Generation (RAG) Implementation}
\label{subsubsec:rag}
Our RAG implementation addresses LLM hallucination by retrieving real-time PyPI metadata (550,000+ modules). However, raw metadata can contain considerable unwanted information, that can mislead models. To address this, we filter the metadata to within the release window of the Python version being used, as well as the explicit Python version support declared by the module. This ensures that only relevant versions are considered, reducing noise and improving inference accuracy.
Filtered metadata is provided as comma-separated version lists, which proved most effective for LLM prompting. Furthermore, in situations where no suitable versions are found, we fall back to the latest release of the module.

\subsection{Parallel Multi-Version Execution}
\label{subsec:parallel}
A distinctive feature of \approach{} is its capacity to validate against multiple Python versions simultaneously. For example, if the LLM predicts Python 3.5 and we have allowed for a range of 2, we execute against versions 2.7, 3.4, 3.5, 3.6, and 3.7 in parallel, increasing success likelihood across deployment scenarios. 

\section{Experimental Setup}\label{sec:studydesign}
The \textit{goal} of this study is to evaluate the effectiveness of \approach{} in resolving Python dependency conflicts for real-world Python programs. To this aim, we formulated the following research questions:
\begin{enumerate}
    \item[\textbf{RQ1}:] \textit{To what extent can current LLMs infer working dependencies from a given Python file?}
    \item[\textbf{RQ2}:] \textit{How does \approach{} compare to state-of-the-art knowledge-based dependency resolution techniques?}
    \item[\textbf{RQ3}:] \textit{Under what conditions does \approach{} outperform traditional methods?}
\end{enumerate}

\textbf{RQ1} evaluates current LLM capabilities for dependency inference. \textbf{RQ2} and \textbf{RQ3} compare \approach{} against knowledge-based approaches to determine contexts where \approach{} excels while identifying potential limitations.

\subsection{LLM Model Selection}\label{subsec:modelselection}
To ensure openness and reproducibility in our approach, we evaluated six open-source LLMs via Ollama\footnote{\url{https://ollama.com}}: three \texttt{Gemma2}~\cite{gemmateam2024gemma2improvingopen} versions (3B, 9B, 27B), \texttt{DeepSeek-R1}~\cite{deepseekai2025deepseekr1incentivizingreasoningcapability}, \texttt{Llama3.1}~\cite{grattafiori2024llama3herdmodels}, and \texttt{MistralAI}~\cite{jiang2023mistral7b}.
\subsection{Dataset}
We validated our approach using the \hgdataset{} dataset, which has become the standard benchmark for evaluating Python dependency resolution tools~\cite{cheng2024readpy, yepyego2022}. This dataset, originally created as part of the \texttt{DockerizeMe} paper by Horton and Parnin~\cite{horton2019dockerizemeautomaticinferenceenvironment}, is a curated subset of the broader Gistable corpus~\cite{gistable2018horton}. It contains $2,891$ "hard" Gists—-Python programs collected from GitHub's Gist platform--that still fail with \texttt{ImportError} even after applying Gistable's naïve resolution strategy. As such, it represents realistic and challenging dependency drift scenarios in Python.

To answer \textbf{RQ1}, we created a subset of the \hgdataset{} to reduce computational cost while preserving result validity. The subset, containing 422 randomly selected Gists, was executed five times per LLM with different seeds to account for stochasticity~\cite{sallou2024breaking}. By using power analysis~\cite{triola2004elementary}, we yield a confidence level of 95\% and a margin of error below 4.5\% for binary outcomes. For \textbf{RQ2} and \textbf{RQ3}, we evaluated the best \approach{} configuration (as identified in \textbf{RQ1}) and the baselines on the full \hgdataset{} dataset to ensure a comprehensive comparison.

\subsection{Implementation and Parameter settings}
We implemented \approach{} using Python 3.11, Langchain 0.2.1, and Ollama 0.2.0. Experiments ran in docker containers on an an AMD EPYC 7713 64-core Processor running at 2.6 GHz (256 CPUs) with an available Nvidia A40 GPU (48GB GDDR6 memory).

Our implementation supports a set of configurable parameters to optimize evaluation and test execution. The key parameters relevant to this study include: \texttt{--model=gemma2}, \texttt{--temp=0.7}, \texttt{--loop=10}, \texttt{--range=1}, and \texttt{--rag=True}.

The \texttt{model} parameter specifies which LLM backend to use. This can be set to any model available in the Ollama model library\footnote{\url{https://ollama.com/library}}. For this study, we primarily used \texttt{Gemma2}, though other models were evaluated as discussed in Section~\ref{subsec:modelselection}.

The \texttt{temp} parameter controls the temperature for the LLM, with higher values (up to 1.0) encouraging more creative responses. We set this to 0.7 to strike a balance between determinism and variation in output. While the default temperature is not standardized—OpenAI’s GPT-4 technical report~\cite{openai2024gpt4technicalreport} uses 0.6 arbitrarily, and Ollama’s documentation\footnote{\url{https://github.com/ollama/ollama/blob/main/docs/modelfile.md}} lists both 0.7 and 0.8 as defaults—we selected 0.7 based on its initial status as Ollama’s default during early development of \approach{}.

As described in Section~\ref{subsec:validation}, \approach{} uses an iterative loop to refine dependency suggestions. We set the \texttt{loop} value to 10 based on empirical trade-offs between runtime and success rate. Lower values led to incomplete resolutions, while higher values increased runtime with diminishing returns.

The \texttt{range} parameter defines how many Python versions are validated in parallel (see Section~\ref{subsec:parallel}). A value of 1 triggers validation across three versions (e.g., $v{-}1$, $v$, $v{+}1$), offering both breadth and efficiency.

Lastly, the \texttt{rag} flag controls whether Retrieval-Augmented Generation is enabled. It is set to \texttt{True} by default but was explicitly disabled when evaluating \textbf{RQ1} to isolate the performance of the LLMs without external metadata assistance.

\subsubsection{\textit{Baseline settings}}\label{subsub:baselinesettings}
We evaluated \approach{} against two state-of-the-art baselines introduced in prior work: \baseline{}~\cite{yepyego2022} and \baselinetwo{}~\cite{cheng2024readpy} (see Section~\ref{sec:related}). Both baselines were executed using the default settings recommended in their respective repositories. Each required Docker-based execution and the creation of a Neo4j knowledge graph. \baseline{} provided a database dump in its repository, while \baselinetwo{} offered this as a separate artifact. Two knowledge graphs are available in \baselinetwo{}; for consistency, we selected \textbf{KG0}, which focuses solely on Python module inference. This aligns with the scope of \approach{}, which does not attempt to resolve OS-level dependencies. It is worth noting that \baseline{} does fix both Python and system-level dependencies, but for evaluation purposes, we restrict comparisons to Python modules only.

\subsection{Evaluation Criteria}\label{ref:subevalcriteria}
To answer \textbf{RQ1}, we performed a comparative analysis of various LLMs to understand their performance within the \approach{} workflow. As part of this analysis, we executed these models with RAG enabled and disabled to understand the effects of RAG in the pipeline. We executed each model a total five times against a subset of the \hgdataset{} dataset, calculating the average number of fixes discovered.

The top-performing model from \textbf{RQ1} was then used in \textbf{RQ2} to compare \approach{} against the two baselines, \baseline{} and \baselinetwo{}, across the full \hgdataset{} dataset. Following the evaluation criteria established in DockerizeMe~\cite{horton2019dockerizemeautomaticinferenceenvironment}, we define a successful fix as a Python program that executes without critical runtime errors such as \texttt{ImportError}, \texttt{Module}-\texttt{NotFoundError}, \texttt{AttributeError}, or \texttt{SyntaxError}.

For each approach, we also recorded the time taken to infer a working environment, as we were particularly interested in understanding the efficiency of each method in resolving dependency conflicts. Accurate timing information provides insight into the practical usability of these approaches, particularly in scenarios where dependency resolution speed is crucial. Some adjustments were required to obtain timings for our baselines. For \baselinetwo{}, more extensive modifications were necessary to achieve precise timing measurements. First, we recorded the inference time, which includes the stage where a Dockerfile was created with the inferred Python modules. Next, we validated the Dockerfiles, as no logging was initially provided to confirm whether a Dockerfile was executable. Each Dockerfile was built and run at this stage, analyzing the output and logging the time required for completion. We then combined these timings to determine which Dockerfiles were successfully executed.
For \baseline{}, we could infer considerable information from their output logs, which contained various data points, including the start and stop times for each program inference and whether the inference was successful.
To account for the stochastic nature of LLMs, we executed \approach{} 10 times, aggregating the data to understand the full extent of fixes. However, as both of our baselines utilize a static knowledge graph, these approaches only required a single run.

To answer \textbf{RQ3}, we conducted a permutation test~\cite{crawley2014statistics}, a non-parametric alternative to the ANOVA test~\cite{st1989analysis}, that does not require the data to be normally distributed. 
This test allowed us to determine whether there is a significant difference in the number of successful fixes produced by \approach{}, \baseline{}, and \baselinetwo{}. We also assessed the impact of various project co-factors on these outcomes. 
The permutation test provided insights into the statistical significance of differences in successful fixes among the approaches without relying on the normality assumptions of ANOVA. To further investigate specific patterns, we analyzed the frequency of successful fixes for particular types of modules (e.g., PyTorch) and for projects with varying numbers of dependencies. We examined the success frequencies across Python files with similar dependencies and different dependency counts. Additionally, we analyzed the intersection and unique sets of projects successfully fixed by \approach{}, \baseline{}, and \baselinetwo{}, identifying both shared and unique successes for each approach.

The complete replication package, including the datasets, the source code, and results, is available on Figshare~\footnote{\url{10.6084/m9.figshare.29204693}}.

\section{Empirical Results}\label{sec:empirical}
In this section, we discuss the results of the experiment with respect to our research questions.

\subsection{Results for RQ1}\label{sec:resultsrq1}

\begin{table}[t]
\caption{Total number of successful fixes for each of the models with RAG on and off. \texttt{(G2-S:Gemma2-2B, G2:Gemma2-9B, G2-L:Gemma2-27B, DS:DeepSeek-R1-8B, L3.1:Llama3.1-8B, M:MistralAI)}}
\label{tab:modeltest}
\centering
\begin{tabular}{c ccccccc}
    \toprule
    RAG & G2-S & G2 & G2-L & DS & L3.1 & M \\
    \midrule
    Enabled & 86 & \cellcolor{lightgray}203 & 187 & 166 & 197 & 194 \\
    Disabled & - & 171 & \cellcolor{lightgray}207 & 164 & 111 & 106 \\
\bottomrule
\end{tabular}
\end{table}

\begin{table}[t]
\caption{Distribution of successful fix counts per Gist in the \hgdataset{} subset dataset.}
\label{tab:countdist}
\centering
\begin{tabular}{l ccccccc}
    \toprule
    Model & RAG & 5 & 4 & 3 & 2 & 1 \\
    \midrule
    \texttt{DeepSeek-R1-8B} & \ding{51}  & 80 & 55 & 36 & 36 & 30 \\
    \texttt{Gemma2-2B} & \ding{51}  & 33 & 31 & 21 & 22 & 38 \\
    \texttt{Gemma2-9B} & \ding{51} & \cellcolor{lightgray}166 & 22 & 12 & 19 & 25 \\
    \texttt{Gemma2-27B} & \ding{51}  & 146 & 28 & 17 & 10 & 23 \\
    \texttt{Llama3.1-8B} & \ding{51}  &  103 & 53 & 38 & 39 & 35 \\
    \texttt{MistralAI} & \ding{51}  & 122 & 49 & 23 & 31 & 35 \\
    \midrule
    \texttt{DeepSeek-R1-8B} & \ding{55} & 75 & 53 & 43 & 36 & 33 \\
    \texttt{Gemma2-2B} & \ding{55} & - & - & - & - & - \\
    \texttt{Gemma2-9B} & \ding{55} & 119 & 33 & 18 & 23 & 27 \\
    \texttt{Gemma2-27B} & \ding{55} & \cellcolor{lightgray}165 & 21 & 19 & 25 & 21 \\
    \texttt{Llama3.1-8B} & \ding{55} & 49 & 25 & 35 & 23 & 61 \\
    \texttt{MistralAI} & \ding{55} & 53 & 24 & 25 & 31 & 33 \\
    
\bottomrule
\end{tabular}
\end{table}

\begin{figure}[t]
\centering
\begin{minipage}[c]{0.48\textwidth}
    \centering
    \includegraphics[width=\linewidth]{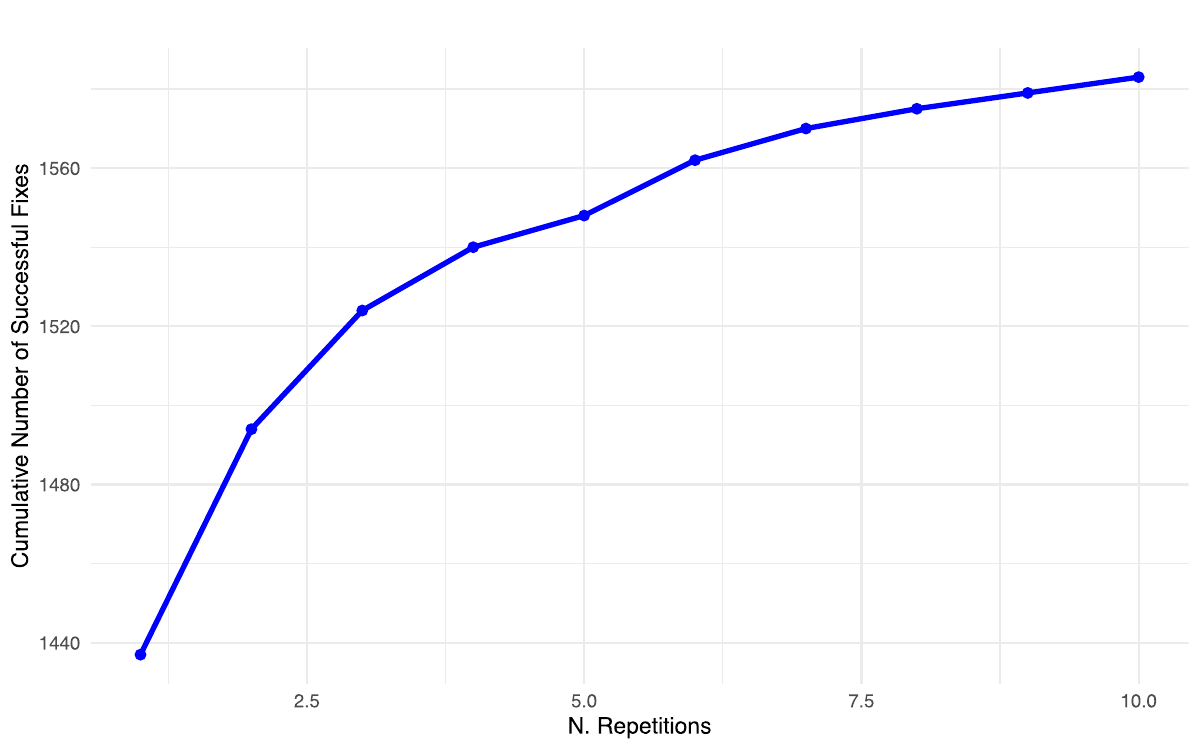}
    \caption{Cumulative chart displaying the number of fixes found by \approach{} across multiple runs.}
    \label{fig:cumulativeplot}
\end{minipage}
\end{figure}

Table~\ref{tab:modeltest} indicates comparable performance across most models ($\sim$200 fixes), except \texttt{Gemma2-2B} which achieved less than half. RAG improved performance for most models, though \texttt{Gemma2-27B} performed slightly better without RAG. Considering both performance and consistency (Table~\ref{tab:countdist}), we selected \texttt{Gemma2-9B} with RAG for subsequent experiments. This model provided an optimal balance between high success rates and computational efficiency, showing consistent performance across multiple runs, feasible for extensive evaluation.

Figure~\ref{fig:cumulativeplot} shows the cumulative fixes for the \texttt{Gemma2-9B} model against the entire \hgdataset{}, over ten independent runs. A single run achieves $\sim$1,440 fixes, with additional runs uncovering more unique solutions due to LLM stochasticity. Even with a single run, \approach{} outperforms both baselines (see more details in the next subsection).

\begin{table}[t]
\caption{Number successful and unsuccessful fixes produced by \approach{}, \baseline{} and \baselinetwo{}. We also report the average time (in seconds) needed to generate the successful fixes alongside the interquartile range (IQR).}
\label{fig:successtable}
\centering
    \begin{tabular}[t]{lrrc}
    \toprule
    Method & \# Fixed & \# Unfixed & Fix Time (IQR)\\
    \midrule
    \approach{} & 1583 & 1308 & 151.461 (221.30)\\
    \baselinetwo{} & 1365 & 1526 & 62.77 (40.57)\\
    \baseline{} & 1302 & 1589 & 4.025 (7.56)\\
\bottomrule
\end{tabular}
\end{table}

\begin{figure}[t]
\centering
\begin{minipage}[c]{0.33\textwidth}
    \includegraphics[width=\linewidth]{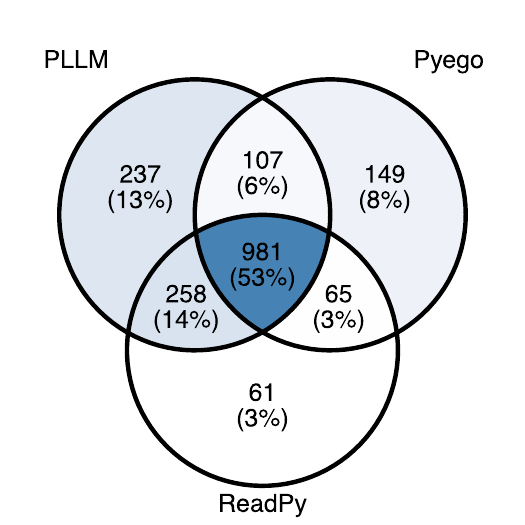}
    \caption{Intersections and difference sets for the projects successfully fixed by \approach{}, \baseline{}, and \baselinetwo{}.}
    \label{fig:venn-diagram}
\end{minipage}
\end{figure}

\subsection{Results for RQ2}\label{sec:resultsrq2}
Table~\ref{fig:successtable} reports the number of successful and unsuccessful fixes produced by \approach{}, \baseline{}, and \baselinetwo{},
and the average time (in seconds) required to generate successful fixes, along with the interquartile range (IQR). As observed, \approach{} outperforms both baselines in terms of the number of successful fixes, achieving a total of $1583$ successful fixes. This is significantly higher than the $1365$ and $1302$ successful fixes produced by \baselinetwo{} and \baseline{}, respectively. 

Compared to the baselines, average fix time is higher for \approach{},
with an average time of 151.46 seconds (compared to 62.77 seconds for \baselinetwo{} and 4.025 seconds for \baseline{}). The higher IQR for \approach{} indicates a broader range of fix times, largely due to the iterative nature of our approach, which continues searching until a successful set of dependencies is found. We argue that this additional time is justified by the higher success rate achieved by \approach{}, and an average time of two minutes remains reasonable for achieving a successful fix with Dockerfile generation and execution.


To further analyze the effectiveness of our approach, we examined cases where \approach{} succeeded while both baselines failed. Figure~\ref{fig:venn-diagram} visualizes the overlap of successful fixes across all three approaches, illustrating the unique effectiveness of \approach{}.
Specifically, all three approaches were successful on 53\% of all Gists, while \approach{} uniquely succeeded on 13\% (representing $237$ Gists) that neither baseline could fix.
There is a distinct overlap of $981$ Gists, representing cases where all three approaches were successful. The combined success across all approaches reaches $1858$ Gists. Considering these $1858$ as the total potentially fixable Gists within our validation system, \approach{} successfully fixed 85\% of the fixable Gists in the dataset. This finding strongly supports the effectiveness of our approach in reliably inferring working dependencies from a Python file and highlights the potential of hybrid approaches, which are part of our future agenda.

\begin{table}[t]
\caption{Percentage of successful fixes produced by \approach{} and \baseline{} for projects with a different number of dependencies. The best values are highlighted in gray color}
\label{tab:interaction}
\centering
\begin{tabular}{c ccc}
\toprule
\# Project Dependencies & \approach{} & \baseline{} & \baselinetwo{} \\
\midrule
0-2 & \cellcolor{lightgray}0.5478 & 0.4846 & 0.5046 \\
3-4 & \cellcolor{lightgray}0.5500 & 0.4162 & 0.4517 \\
5-6 & \cellcolor{lightgray}0.5410 & 0.2551 & 0.2806 \\
7+ & \cellcolor{lightgray}0.5333 & 0.3555 & 0.2222 \\
\bottomrule
\end{tabular}
\end{table}

\begin{table}
\caption{Percentage of successful fixes for the top-15 most frequent Python modules in our dataset for \approach{}, \baseline{}, and \baselinetwo{}.}
\label{tab:dependencies}
\centering
\small
\begin{tabular}[t]{l rrrr}
\toprule
\textbf{Module Name} & \textbf{\# Projects} & \approach{} & \baseline{} & \baselinetwo{}\\
\midrule
numpy & 572 & \cellcolor{lightgray}69.06 & 57.17 & 53.50\\
django & 382 & \cellcolor{lightgray}86.65 & 65.18 & 79.84\\
scipy & 296 & \cellcolor{lightgray}79.39 & 58.78 & 62.50\\
requests & 168 & \cellcolor{lightgray}66.67 & 50.60 & 52.38\\
pillow & 166 & \cellcolor{lightgray}76.51 & 69.88 & 71.08\\
matplotlib & 162 & \cellcolor{lightgray}64.81 & 62.96 & 54.94\\
tensorflow & 140 & \cellcolor{lightgray}83.57 & 51.43 & 62.14\\
scikit-learn & 136 & \cellcolor{lightgray}76.47 & 55.88 & 66.91\\
tensorflow-gpu & 135 & \cellcolor{lightgray}82.96 & 52.59 & 68.15\\
opencv-python & 85 & \cellcolor{lightgray}64.71 & 49.41 & 24.71\\
pandas & 73 & \cellcolor{lightgray}68.49 & 63.01 & 57.53\\
cython & 69 & \cellcolor{lightgray}55.07 & 33.33 & 39.13\\
image & 69 & \cellcolor{lightgray}23.19 & 18.84 & 18.84\\
pyyaml & 69 & 66.67 & 66.67 & \cellcolor{lightgray}71.01\\
theano & 67 & \cellcolor{lightgray}68.66 & 26.87 & 37.31\\
keras & 63 & \cellcolor{lightgray}77.78 & 14.29 & 42.86\\
\bottomrule
\end{tabular}
\end{table}

\subsection{Results for RQ3}\label{sec:resultsrq3}
Permutation tests confirm \approach{}'s statistical superiority ($p<0.01$) and significant interaction with dependency count. Table~\ref{tab:interaction} shows \approach{} maintains consistent success (53-55\%) regardless of dependency count, while baselines degrade from ~50\% to 32\% (\baseline{}) and $<$25\% (\baselinetwo{}), indicating knowledge-graph methods struggle with complex projects.

\begin{table*}[t]
\caption{Example Gists showing where \approach{}, \baseline{}, and \baselinetwo{} succeeded or failed in dependency resolution. Each cell lists inferred dependencies; \cellcolor{lightgray}grey cells denote successful runs. These cases illustrate complementary strengths and limitations of the different approaches.
}
\label{tab:snippetsdetails}
\centering
\small
\begin{tabular}{c ccc}
\toprule
Gist ID & \approach{} & \baseline{} & \baselinetwo{} \\
\hline
019fd5c706e0bc94879f & \cellcolor{lightgray} rx;twisted & rx;twisted & twisted;urx \\
\hline
477a9dcd198439ef2def & urllib2;readability & \cellcolor{lightgray} readability-lxml & \cellcolor{lightgray} readability-lxml\\
\hline
3153844 & \cellcolor{lightgray} graphite& NA  & graphiti \\
\hline
4e5035242b8e4b07ff3a & pymongo;pip;bson  & 
\cellcolor{lightgray} pymongo & \cellcolor{lightgray} pymongo \\
\hline
c2dfe5772ba3cd16c1be17ba42b7db66 & \cellcolor{lightgray} keras;tensorflow & keras  & keras;tensorflow-gpu\\
\hline
7030355 & mpd & \cellcolor{lightgray}python-mpd2  & \cellcolor{lightgray} python-mpd2\\
\bottomrule
\end{tabular}
\end{table*}

Table~\ref{tab:snippetsdetails} displays examples of Gists in which the various approaches passed and failed. We focus here on examples where \approach{} was either better or worse than both baselines (\eg~\approach{} passed, \baseline{}, \baselinetwo{} failed, \etc). This allows us to dig deeper into the results of these specific data points to understand why the LLM succeeds or fails.

In Gist \texttt{019fd5c706e0bc94879f}, \approach{} and \baseline{} both identified \texttt{rx} and \texttt{twisted} as required modules. In contrast, \baselinetwo{} incorrectly inferred \texttt{urx}, likely due to a semantic match in its knowledge graph. This resulted in an \texttt{ImportError} and failure. While \baseline{} selected the correct modules, its choice of \texttt{rx} version also failed at runtime. All three approaches selected the same version of \texttt{twisted}. \approach{} required a few iterations to converge on a compatible version of \texttt{rx}, ultimately producing a successful configuration.

Gist \texttt{`477a9dcd198439ef2def'} is an example of where \approach{} was unsuccessful. By observing the dependency chosen by both baselines, we identify that \texttt{readability} is likely no longer available or has been replaced with \texttt{readability-lxml}. We can also observe the same scenario with \texttt{`7030355'} and the \texttt{mpd}, now \texttt{python-mpd2} dependency. These examples could be added to our JSON file mentioned in Section~\ref{subsec:stageb}. We also note a failure for \approach{} with Gist \texttt{`4e5035242b8e4b07ff3a'}. This is a little more interesting as the LLM has \texttt{pymongo} listed as a dependency. However, unlike the other approaches, it has also attempted to install \texttt{pip} and \texttt{bson}, which have likely contributed to the failure in this instance.

Finally, we examine Gists \texttt{`3153844'} and \texttt{`c2dfe57}-\texttt{72ba3cd16c1be17ba42b7db66'} in which \approach{} was the only successful approach. Here, we witness another example of \baselinetwo{} attempting a dependency with an alternative name. In this instance, the original dependency \texttt{graphite} was successfully installed by \approach{}, and no dependency was even attempted by \baseline{}. The final Gist \texttt{`c2dfe5772ba3cd16c1be17ba42b7db66'} has overlap between the baselines, but ultimately, the lack of \texttt{tensorflow} in \baseline{} and the incorrect \texttt{tensorflow} dependency for \baselinetwo{} led to unsuccessful conflict resolutions for both approaches.

These examples illustrate several core strengths of \approach{}: iterative refinement, flexible version search, and responsiveness to real-time failure signals. However, they also expose failure modes tied to naming mismatches and unnecessary package inference—areas where targeted prompt tuning or expanded mapping rules could yield further gains.

\section{Threats to validity}\label{sec:threats}
\textit{Construct validity.}
A key aspect of our approach (\approach{}) is allowing the LLM to dynamically infer Python versions from Python code and validate these along with adjacent Python versions. This flexible strategy addresses limitations in prior work~\cite{cheng2024readpy}, which rely on static Python versions during testing, potentially limiting both the adaptability and real-world relevance of the results. The only exception in our approach is Python 2.7, which is consistently validated due to its unique, legacy-specific versioning requirements.

Our method ensures that, once execution begins, the Python version remains unchanged, even if a \texttt{SyntaxError} occurs ---an indicator of an incompatible Python version. Unlike approaches that dynamically switch versions mid-execution, our approach simulates real-world constraints more accurately. 

\textit{Internal validity.}
For evaluation, we used a single LLM, Gemma2~\cite{gemmateam2024gemma2improvingopen}, which showed strong performance in tests compared to other open-source LLMs. A small manual study was performed using the commercial Claude 4 Sonnet model~\footnote{\url{https://www.anthropic.com/claude/sonnet}}. Claude fixed 15 PLLM-failure cases, already handled by the baselines, but failed on all PLLM-only successes particularly with 3+ dependencies, often hallucinating versions or misusing import names. This supports our open-source choice.

To address the stochastic nature of LLM outputs, we followed existing guidelines~\cite{sallou2024breaking} on assessing LLMs and executed \approach{} multiple times, aggregating the results to ensure a more comprehensive assessment of its effectiveness. The model size also impacts internal validity; we selected the 9-billion-parameter version of Gemma2, compatible with both our local machine and the validation servers.

\textit{External validity.}
Our evaluation is based on the \hgdataset{} dataset, which has become a de facto standard in the Python dependency resolution literature. It has been used by both of our baselines~\cite{yepyego2022, cheng2024readpy}, and contains realistic, challenging Python programs (Gists) exhibiting dependency drift, making it well suited for assessing module-level inference capabilities.

While \baselinetwo{} has also been evaluated on other datasets, we focused on \hgdataset{} for two key reasons. First, it enables direct, fair comparison with prior work. Second, it allows us to isolate and evaluate the core question of this paper: how well can LLMs infer and resolve dependency configurations from Python source code? This controlled scope provides a focused evaluation of dependency inference capabilities, which is fundamental to broader dependency resolution challenges.
Expanding to other datasets would improve the generalizability of our findings, and this is a natural next step for future work.

\textit{Conclusion validity.}
The \hgdataset{} dataset includes some Python programs that are unresolvable due to module deprecation. Additionally, our approach was validated only on x86 Docker Linux machines. This limitation restricts our ability to validate Gists in \hgdataset{} designed for other platforms, such as macOS or ARM-based systems. Hence, we were unable to validate if a Dockerfile generated by one of our baselines would execute correctly on these platforms.

As a result, \baseline{} includes operating system-level dependencies by default. \baselinetwo{} also provides this capability through a specific database dump. However, to keep the approaches consistent with \approach{}, we configured \baselinetwo{} to only fix Python-level dependencies with its default database.

\section{Conclusion and Future Work}
\label{sec:conclusion}

This paper introduced \approach{}, a novel LLM-driven approach for resolving Python dependency conflicts. Unlike traditional methods such as \baseline{} and \baselinetwo{}, which rely on precomputed knowledge graphs, \approach{} uses prompt-based inference and feedback loops to recover a working environment from raw Python code and runtime errors. This eliminates the need for extensive external infrastructure, making the solution lightweight, adaptable, and easy to apply across environments.

Our evaluation on the \hgdataset{} benchmark demonstrates that \approach{} outperforms the state-of-the-art solutions \baseline{}{} and \baselinetwo{}. \approach{} achieves significantly more successful resolutions, performing consistently, even in cases with many dependencies. Qualitative analysis reveals that \approach{}'s success stems from its ability to refine its guesses based on runtime information ---an advantage absent in static systems.

At the same time, both baselines still succeed in a subset of cases that \approach{} fails to resolve. These are often linked to legacy dependencies, renamed packages, or system-level nuances more effectively handled by structured knowledge graphs. This points to a key insight: prompt-based and knowledge-based techniques offer complementary strengths. Rather than treating them as competing approaches, we see clear potential in hybrid approaches that unify LLM-driven reasoning with curated symbolic knowledge.

Future work includes extending \approach{} to support system-level dependency resolution, improving prompt construction and dependency disambiguation, and exploring alternative RAG methods such as \texttt{GraphRAG}~\cite{graphrag2024}. Finally, we plan to investigate hybrid architectures that combine prompt-based inference with knowledge graph approaches given their complementary nature as highlighted by our results.

\bibliographystyle{IEEEtran}
\bibliography{references}

\end{document}